\documentclass[10pt]{article}
\usepackage{geometry} 
\usepackage{authblk}
\usepackage{graphicx}
\usepackage{epsfig}
\usepackage{xcolor}
\usepackage{ulem}
\usepackage{float}
\usepackage{soul}
\usepackage{lineno}
\newcommand{\URuSi}{URu$_2$Si$_2$}

\title{Topological Excitations of Hidden Order in \URuSi\ Under Extreme Electric Fields}
\author[1]{Laurel E. Winter\thanks{Corresponding Author: lwinter@lanl.gov}}
\author[2]{Arkady Shekhter}
\author[3]{Brad Ramshaw}
\author[4]{Ryan E. Baumbach\thanks{Present address: Condensed Matter Group, National High Magnetic Field Laboratory, Florida State University, Tallahassee Florida, 32310, USA }}
\author[4]{Eric D. Bauer}
\author[1]{Neil Harrison}
\author[5,6]{Philip J. W. Moll}
\author[1]{Ross D. McDonald}
\affil[1]{National High Magnetic Field Laboratory, Los Alamos, New Mexico 87544, USA}
\affil[2]{National High Magnetic Field Laboratory, Florida State University, Tallahassee, Florida 32310, USA}
\affil[3]{Laboratory of Atomic and Solid State Physics, Cornell University, Ithaca, New York 14853, USA}
\affil[4]{Los Alamos National Laboratory, Los Alamos, New Mexico 87544, USA}
\affil[5]{Max Planck Institute for Chemical Physics of Solids, Dresden, Germany D-01187}
\affil[6]{Institute of Materials, EPFL, 1015 Lausanne, Switzerland}

\date{} 

\begin{document}

\maketitle{}

{\bf Quantum materials are epitomized by the influence of collective modes upon their macroscopic properties. Relatively few examples exist, however, whereby coherence of the ground-state wavefunction directly contributes to the conductivity. Notable examples include the quantizing effects of high magnetic fields upon the 2D electron gas, the collective sliding of charge density waves subject to high electric fields, and perhaps most notably the macroscopic phase coherence that enables superconductors to carry dissipationless currents. Here we reveal that the low temperature hidden order state of \URuSi\ exhibits just such a connection between the quantum and macroscopic worlds -- under large voltage bias we observe non-linear contributions to the conductivity that are directly analogous to the manifestation of phase slips in one-dimensional superconductors \cite{Little67}, suggesting a complex order parameter for hidden order.}

For over 30 years the intermetallic heavy-fermion \URuSi\ has remained at the forefront of the correlated electron problem due in part to the long-standing mystery surrounding the unknown broken symmetry associated with the hidden order state that exists below  T$_{\rm HO} =$ 17.5~K \cite{Palstra85, Maple86, Schlabitz86, Mydosh11}.  Attempts to resolve the hidden order conundrum broadly fall into two classes of experiments.   In the first class are symmetry resolving probes that have primarily focused on neutron scattering \cite{Bourdarot14, Das13} and more recently x-ray \cite{Tonegawa14} and Raman spectroscopy \cite{Kung15}.  In the second class are experiments that explore the thermodynamic response of hidden order, leading to the examination of the rich neighborhood of competing phases through the application of high magnetic fields \cite{Sugiyama90, Ohkuni99, Scheerer12, Harrison13} and pressure \cite{Amitsuka07, Hassinger10, Butch10}, as well as doping and isovalent substitution series \cite{Amitsuka88, Bauer05, Das15, Gallagher15}.  

As hidden order continues to defy clear identification we explore a new approach of probing the non-equilibrium state of the hidden order, allowing one to study the dynamics that are hidden in equilibrium experiments.  Here we focus on applying a strong electric field to \URuSi, driving the conductivity in the hidden order state beyond the linear regime, an experimental approach that has been pivotal to understanding other strongly correlated states such as vortex motion in superconductivity \cite{Kim65, Tinkhambk} and spin- and charge-density waves \cite{Grunerbk}.  In the latter case, the non-linear electric field response of density waves, i.e. density wave sliding, is a key experiment that can distinguish between a localized correlated density wave state and a trivial structural transition to a single-particle insulator.

Until now, the application of an electric field has not been successfully used to study the behavior of the \textit{f}-electrons in the hidden order state of \URuSi \cite{Hundley}, largely due to the significant Joule heating associated with the induced currents that can occur when applying a large electric field across a metal.  Furthermore, making electrical contacts to exotic metals usually results in contact resistances that are the dominant source of dissipation, even when they are only a few ohms.  To mitigate such thermal effects we have employed Focused Ion Beam (FIB) lithography to tailor the geometry of \URuSi\ to produce samples with much larger resistances than otherwise obtainable in small as-grown crystals  \cite{Moll18}; by ion milling long thin current paths on micron-scales, the sample -- not the contacts -- are the dominant resistance, enabling the application of unprecedented electric fields to a metal at cryogenic temperatures.

\begin{figure}
   \centering
   \includegraphics[scale=0.9]{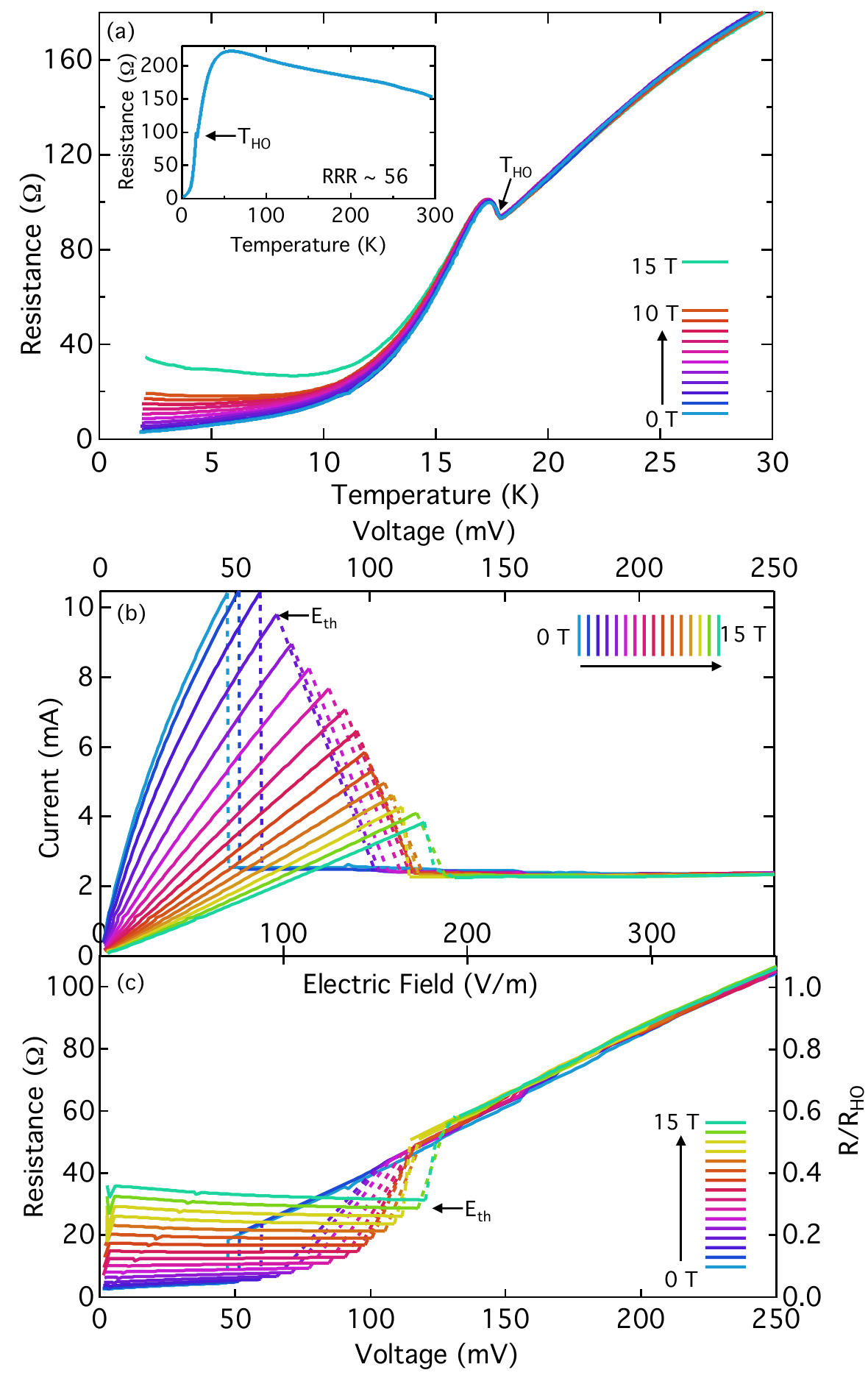}
    \caption{\textbf{Electric transport properties of microstructured \URuSi\ at magnetic fields between 0 and 15 T.}  (a) Four-terminal resistance as a function of temperature at constant magnetic fields, with the inset showing the entire temperature dependence at zero magnetic field from which the residual resistivity ratio (RRR) is determined. The hidden order transition is denoted by T$_{\rm{HO}}$.  (b) Current and (c) four-terminal resistance both as a function of the voltage and the resulting electric field, $E = V/d$, where \textit{d} is the effective length of the microstructured sample and E$_{\rm{TH}}$ is the threshold electric field.  In addition, the right axis of figure (c) corresponds to the ratio of resistance versus the resistance at hidden order in this sample. }
   \label{fig:SF}
\end{figure}

The main observation reported here is an abrupt onset of increased dissipation within hidden order under large voltage bias.  Given the ``N"-type behavior of the negative differential conductance (NDC), whereby current is a single valued function of voltage, but voltage is a multivalued function of current, voltage controlled sweeps are necessary to map the NDC. (See Supplementary Information Section I for further results and discussion of current and voltage controlled measurements.)  Consequently, in this work we set a dc voltage bias across the current carrying contacts to study the non-equilibrium state of hidden order.  Figure~\ref{fig:SF}(b) plots the current through a microstructured \URuSi\ sample at 1.8 K in superfluid $^4$He (for magnetic field between 0 and 15 T) as a function of the voltage drop between the non-current carrying contacts (V) and the resulting electric field, $E = V/d$, where \textit{d} is the effective length of the sample.  At a threshold electric field (E$_{\rm th}$) the initial ohmic behavior abruptly gives way to a decrease in current, as the sample dissipation increases as shown in Figure~\ref{fig:SF}(c).  Biased beyond E$_{\rm th}$, the resistance (and current) of the material collapses onto a single curve irrespective of the applied magnetic field.  The lack of magnetic field dependence in this higher resistance state sharply contrasts with the factor of 20 difference in the ohmic response of the hidden order state between 0 and 15 T at zero voltage bias and is reminiscent of the lack of magnetoresistance observed in \URuSi\ at temperatures above T$_{\rm HO}$ (Figure~\ref{fig:SF}(a)).  It would be natural to speculate that this transition corresponds to Joule heating through T$_{\rm HO}$, yet our sample self-consistently excludes this possibility.  The absolute resistance value immediately above the threshold electric field is approximately half of the resistance at T$_{\rm HO}$ -- as highlighted by the right axis in Figure~\ref{fig:SF}(c) -- therefore providing direct evidence of a sample temperature significantly below T$_{\rm HO}$.  

\begin{figure*}
\centering
\includegraphics[scale=0.6]{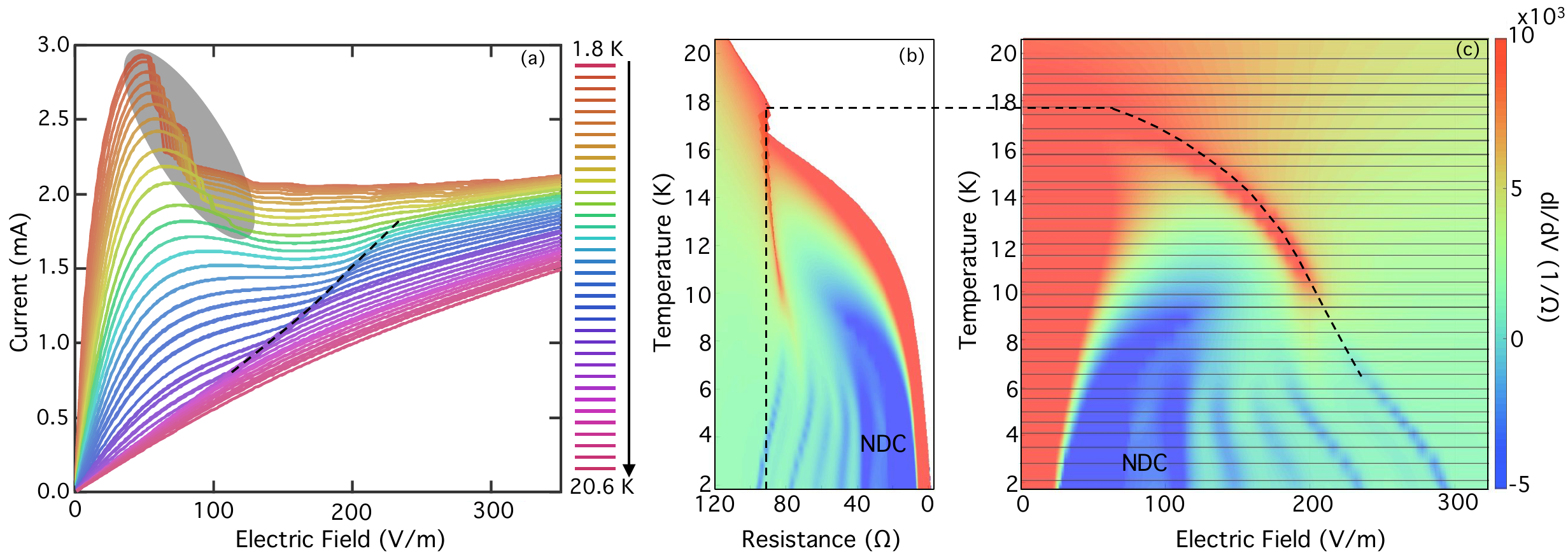}
\caption{\textbf{Temperature dependence of the electric field behavior of microstructured \URuSi\ obtained in $^4$He gas between 1.8 and 20.6 K.}  (a) Current measured across the sample as a function of the resulting electric field (due to applied voltage).  The shaded region and dashed line highlight two features of interest.  (b-c) The calculated differential conductance (\textit{dI/dV}) of the same data shown in the first subfigure, graphed as a function of temperature and resistance or electric field, where red and blue regions highlight the positive differential conductance (PDC) and negative differential conductance (NDC) respectively. The dashed lines correspond to the dashed line feature also shown in the first subfigure.}
\label{fig:IVTemp}
\end{figure*}

The temperature stability of the voltage biased microstructure relies on its thermal contact with the substrate and direct immersion in superfluid $^4$He at 1.8 K.  Once we turn to the temperature dependence of the IV characteristics the cooling power of direct immersion is lost and the effects of self-heating are more evident.  The temperature dependence was studied in $^4$He gas at elevated temperatures, as shown in Figure~\ref{fig:IVTemp}. 
Despite the evidence of Joule heating as indicated by the increase curvature in the nominally ohmic region -- the defining characteristic of negative differential conductance (NDC) -- an E$_{\rm th}$ consistent with the measurements in superfluid helium at 1.8 K is observed.  As highlighted by the shaded region in Figure~\ref{fig:IVTemp}(a) there is only a weak temperature dependence of E$_{\rm th}$.  At higher temperatures a small increase in current is observed at what appears to be a temperature dependent electric field threshold as denoted by the dashed line in the same figure. 
The differential conductance (\textit{dI/dV}), conveyed by the coloring in Figures~\ref{fig:IVTemp}(b,c), shows that this high temperature behavior corresponds to a region of positive differential conductance (PDC, red features).  Furthermore, the PDC emerges from the zero bias resistance peak of the hidden order transition and corresponds with the absolute resistance value at the hidden order transition for this particular sample R$_{\rm HO} = 100\ \Omega$ (Figure~\ref{fig:IVTemp}(b)).  Crucially, this confirms that while self-heating is clearly present in these measurements, the experiment itself can self-consistently distinguish it from the electric field driven increase in dissipation coinciding with the NDC.  
   As a result, the PDC region can be attributed to the thermal suppression of the hidden order due to Joule heating, defining the PDC in Figure~\ref{fig:IVTemp}(c) as an isotherm at T$_{\rm HO} \sim$ 17.5 K, providing an upper temperature bound on the observed NDC (blue features).  This requires that the primary NDC region is not only occurring within the hidden order state, but that it also does not preclude the thermal suppression of hidden order at higher temperatures.   This rules out the drop in current at E$_{\rm th}$ as the suppression of hidden order.
   At electric fields beyond the primary NDC region additional weaker signatures of NDC appear as blue branches at low temperatures in Figures~\ref{fig:IVTemp}(b) and as small steps in Figure~\ref{fig:IVTemp}(c), and are consistent with variations in electric field along the sample due to differences in the effective width of the corners of the meander.
   
 The existence of a threshold electric field provides a means of investigating the non-linear dynamics of the hidden order state in \URuSi.  Since electric field does not naturally provide an intrinsic energy scale, it is necessary to consider the length scale over which it acts.  For single-particle phenomena, such as the NDC exhibited by an Esaki diode \cite{Esaki58} or a superlattice driven into a Bloch oscillator regime \cite{Grondin85, Esaki70}, this scale is provided by the device dimensions parallel to the electric field.  For the hidden order state in \URuSi\ knowledge of the electronic lifetime and dispersion from magneto-quantum oscillation measurements rules out the observed NDC arising from an intrinsic (unit cell) length scale by several order of magnitude (see details in Section III of the Supplementary Information).   This leads to the conclusion that the observed NDC is related to the correlated nature of hidden order.   For many-body phenomena NDC is distinctly ``S" or ``N"-type, depending upon whether the state is intrinsically localized or itinerant.  For the former, the non-linearity is characterized by an increase in conductivity, as displayed by the sliding of a charge density wave (CDW) beyond a threshold electric field when it depins \cite{Grunerbk}.  By contrast an intrinsically conducting condensate exhibits increased dissipation once the electric field is sufficient to discontinuously distort the phase.  In one-dimensional superconductors such phase slips \cite{Michotte03, Meyer75} are evident as the latter ``N"-type NDC.   The IV characteristics of \URuSi\ presented here indicate that the hidden order does not exhibit CDW like pinning, rather behaving as an itinerant condensate.   
 
 \begin{figure}
\includegraphics[scale=0.4]{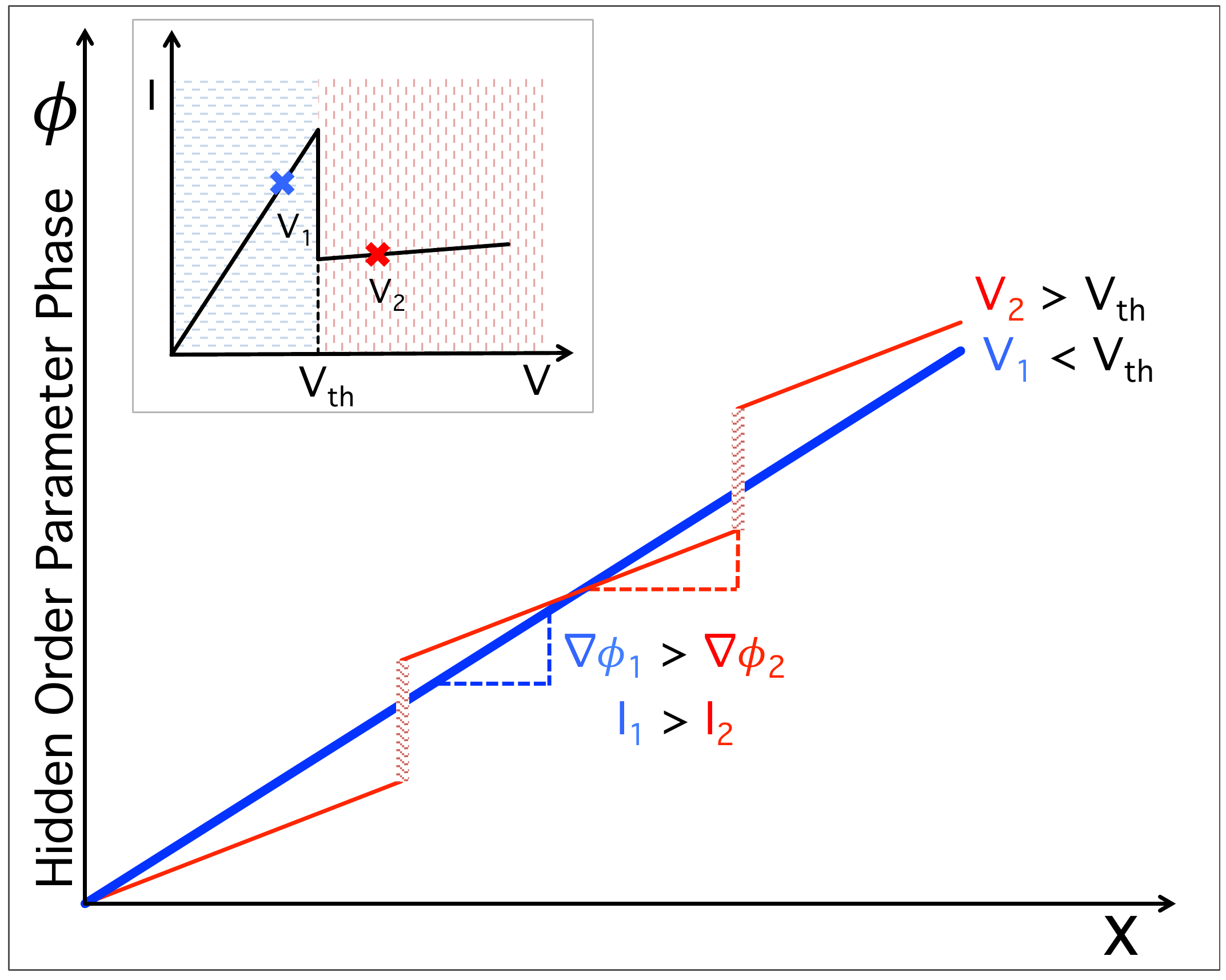}
   \centering
\caption{\textbf{A schematic of a the (hidden) order parameter phase $\phi$ as a function of position $x$ along a sample for a voltage above and below a threshold voltage (V$_{\rm{th}}$).} Below a threshold voltage, V$_1 < ~ $V$_{\rm{th}}$ (thick blue line), $\phi$ is continuous and has a gradient ($\nabla \phi$) proportional to current I$_1$.  For a voltage greater than the threshold, V$_2 > ~ $V$_{\rm{TH}}$ (thin red line), phase slips centers can occur at positions across the sample, which are represented by the patterned vertical lines.  The inset is a schematic of the IV characteristics of a sample undergoing a phase slip in a voltage biased experiment such as presented in this work.}
\label{fig:pic}
\end{figure}

Phase slips  \cite{Anderson66} describe the non-linearity observed in superconductors whose transverse dimensions are smaller than or comparable to their Ginzburg-Landau coherence length, $\xi$ \cite{Skocpol74}. In these contexts superconducting phase slips are topological deformations of the order parameter, $\Psi (r) = |\Psi(r)|e^{i\phi(r)}$, that result in a $\phi = 2\pi$ phase shift leading to a state with lower current and consequently lower free energy.   While nothing in \URuSi\ points towards one-dimensionality, we suggest that the microstructuring of the sample reduces its effective dimensions to one.  Thus, the effective diameter of the crystalline structure is so small that no phase gradient can be supported across the width of the meanders but only along the length.  This scenario suggests that the hidden order has a complex order parameter whose effective coherence length approaches the single meander length of $\sim$ 70 $\mu$m -- consistent with the suggested tens-of-micrometer domain size implied through magnetic anisotropy measurements \cite{Okazaki11}.

If the IV character observed in this work is indeed indicative of phase slips, then it would be the first observation of the behavior in a non-superconductor or in material well above its superconducting transition temperature.   Given this new context in which we have observed evidence for current carried via the gradient of a non-superconducting wavefunction, Figure~\ref{fig:pic} illustrates the spatial variation of phase both above and below the threshold electric field.  For a coherent perturbation of the condensate, current \textit{I} is proportional to the gradient of phase, $\nabla \phi$.  At voltages V$_1 < \rm{V_{th}}$, the sample exhibits ohmic behavior with a continuous evolution of phase along the sample. For a voltage larger than the threshold voltage, V$_2 > \rm{V_{th}}$, the sample contains phase slip centers at various $x$ where the phase discontinuously jumps by 2$\pi$, such that the phase gradient away from the discontinuities decreases ($\nabla \phi_1 > \nabla \phi_2$) resulting in a decrease current at high voltage.  The threshold field E$_{\rm th}$ being determined by a balanced energy cost of these phase distortions and the reduction in current density.

\begin{figure}[h!]
\includegraphics[scale=0.55]{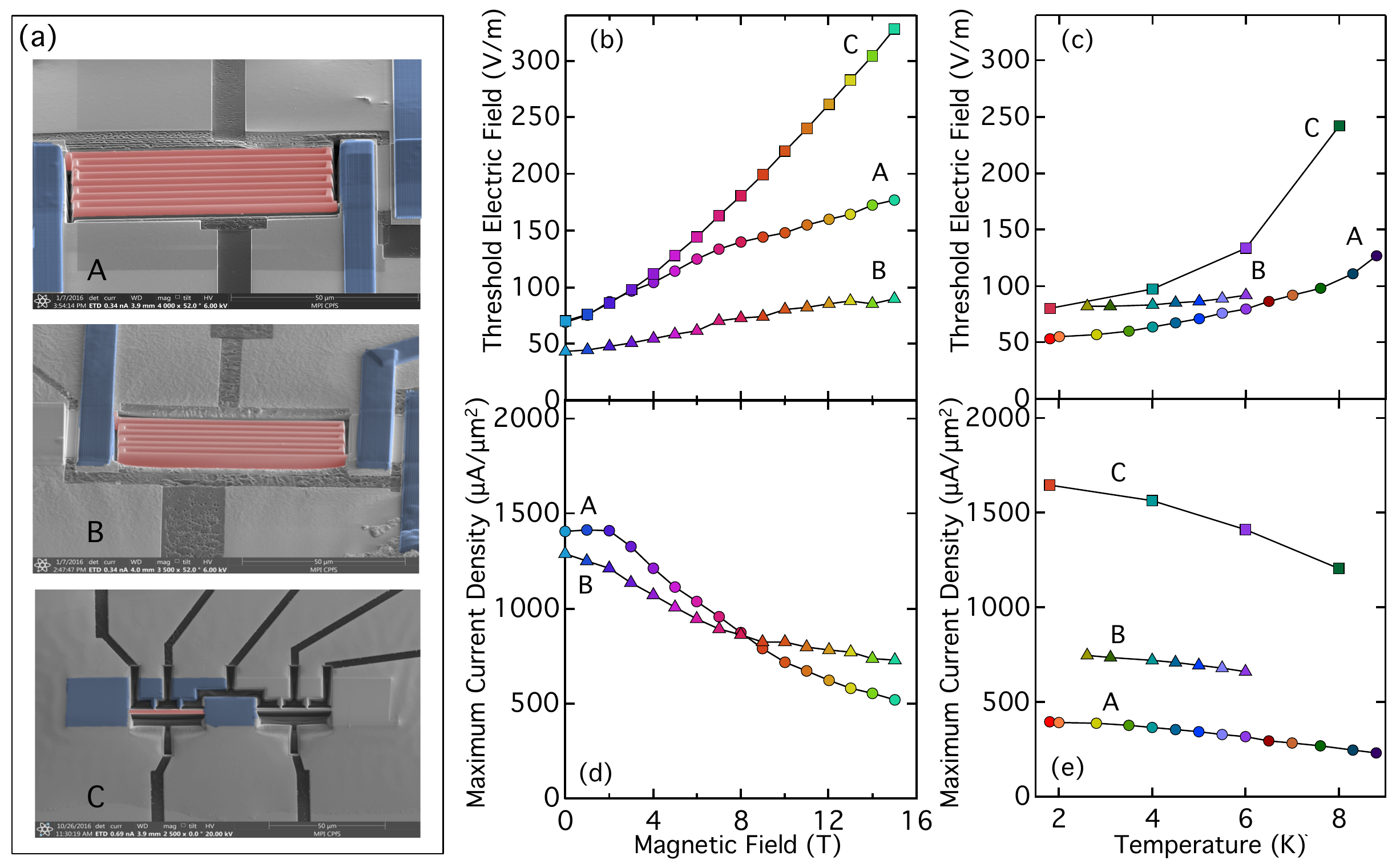}
   \centering
\caption{\textbf{The threshold electric field and maximum current densities as a function of applied magnetic field and temperature for three different microstructured \URuSi\ samples.}  Electron microscope images of the three samples are shown in figures (a), where the microstructure is highlighted in pink and the current and voltage contacts are highlighted in blue.  The threshold electric field and maximum current densities are shown in the remaining figures (b-e), where data for samples A, B, and C are denoted by circles, triangles, and squares respectively, and the colors correspond to the different magnetic fields and temperatures.}
\label{fig:sum}
\end{figure}

To investigate the intrinsic nature of the apparent phase slips, measurements were ultimately carried out on three different microstructured \URuSi\ samples (Figures~\ref{fig:sum}(a)) with varying electrical and geometrical properties as described in Table~\ref{table:table} and in greater detail in Supplementary Information Section II.  Despite the sample variations the threshold electric field values and current density thresholds are comparable for all three samples as a function of magnetic field and temperature as shown in Figures~\ref{fig:sum}(b-e).  To further explore the possibility of phase slips in the hidden order state of \URuSi\ a direct study of the phase coherence would be beneficial.  

\begin{table}[h!]
\centering
\begin{tabular}{|c|c|c|c|c|c|}
\hline
\textbf{Sample} & \textbf{Length} & \textbf{Thickness} & \textbf{R (1.8 K)} &  \textbf{$\rho$ (1.8 K)} & \textbf{RRR}  \\ \hline
A           & 680$\mu$m    & 7.4 $\mu$m &      2.47$\Omega$ & 2.7 $\mu\Omega \cdotp$cm   & 56          \\ \hline
B             & 480 $\mu$m       & 6.8 $\mu$m           & 1.8 $\Omega$  & 2.5 $\mu\Omega \cdotp$cm   &  14            \\ \hline
C             & 11.12$\mu$m        & 5.2 $\mu$m     & 20 m$\Omega$ & 1.6 $\mu\Omega \cdotp$cm  &   115          \\ \hline
\end{tabular}
\caption{\textbf{The effective length and width, as well as the  low-temperature resistance, resistivity, and residual resistivity ratio for three different microstructured \URuSi\ samples.} Sample A is discussed in detail in the text, while results on B and C can be found in Supplementary Information Section II.}
\label{table:table}
\end{table}

\pagebreak
Despite all materials being bound by quantum mechanics only a very limited subset display the macroscopic manifestation of phase coherence that qualify them as quantum materials.
If what we are observing in these microstructured \URuSi\ samples is indeed phase slip behavior, it indicates that hidden order should be considered in such terms \cite{Mermin79}. Specifically since the phase slips are occurring within the hidden order state, it follows that the state's order parameter must (at a minimum) be complex, i.e.\ the system must possess macroscopic (over the micron scale of our devices) coherence of a generalized-phase, the gradient of which carries current. Since there is no evidence that the hidden order is pinned to the lattice, such as in the case of charge density waves, it raises the question as to why this state does not carry a dissipationless supercurrent, drawing analogy to the finite resistance of one-dimensional superconductors.

\subsection{Data Availability}
The data that supports the findings of this study are available from the corresponding author upon reasonable request. 

\subsection{Acknowledgements}
We thank Premala Chandra, Piers Coleman, Nicholas Butch, and Mike Hundley for insightful discussions. L.E.W. acknowledges funding from US DOE through the LANL/LDRD Program and the G. T. Seaborg Institute.  E.D.B acknowledges support from the LANL LDRD project 20170204ER.  N.H. acknowledges DOE Office of Science, BES project ``Science in 100 Tesla" LANLF100.  R.D.M. acknowledges funding from the LANL LDRD DR20160085 'Topology and Strong Correlations'.  A portion of this work was performed at the National High Magnetic Field Laboratory, which is supported by the National Science Foundation Co-operative Agreement No. DMR-1157490 and the State of Florida.  

\subsection{Author Contributions}
R.D.M., N.H., and P.J.W.M. conceived of the study.  E.D.B. and R.E.B. synthesized the crystals. J.W.M. made the micronscale devices.  L.E.W. performed the experiments and analyzed the data with input from R.D.M.  A.S., and B.R. discussed the results and implications.   L.E.W. and R.D.M. wrote the manuscript with input from all of the authors.  

\subsection{Additional Information}

\textbf{Competing Interests:} The authors declare no competing interests.

\pagebreak
\part*{Supplementary Material}

\section{I. Experimental Method}
\begin{figure}[h!]
\includegraphics[scale=0.6]{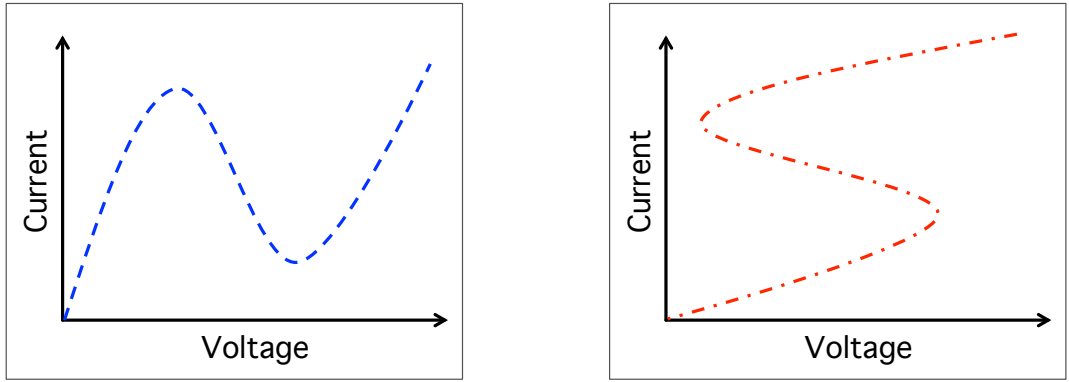}
\centering
\caption{\textbf{Schematics of ``N" and ``S"-type negative differential resistance.} (a) ``N"-type or voltage controlled negative resistance character was observed for microstructured \URuSi\ samples where the sample resistance was equal to or comparable to the contact resistance.  (b) ``S"-type or current controlled negative resistance character was observed in sample C where the contact resistance dominated the sample resistance.}
\label{fig:NandS}
\end{figure}

\begin{figure}[h!]
\includegraphics[scale=0.5]{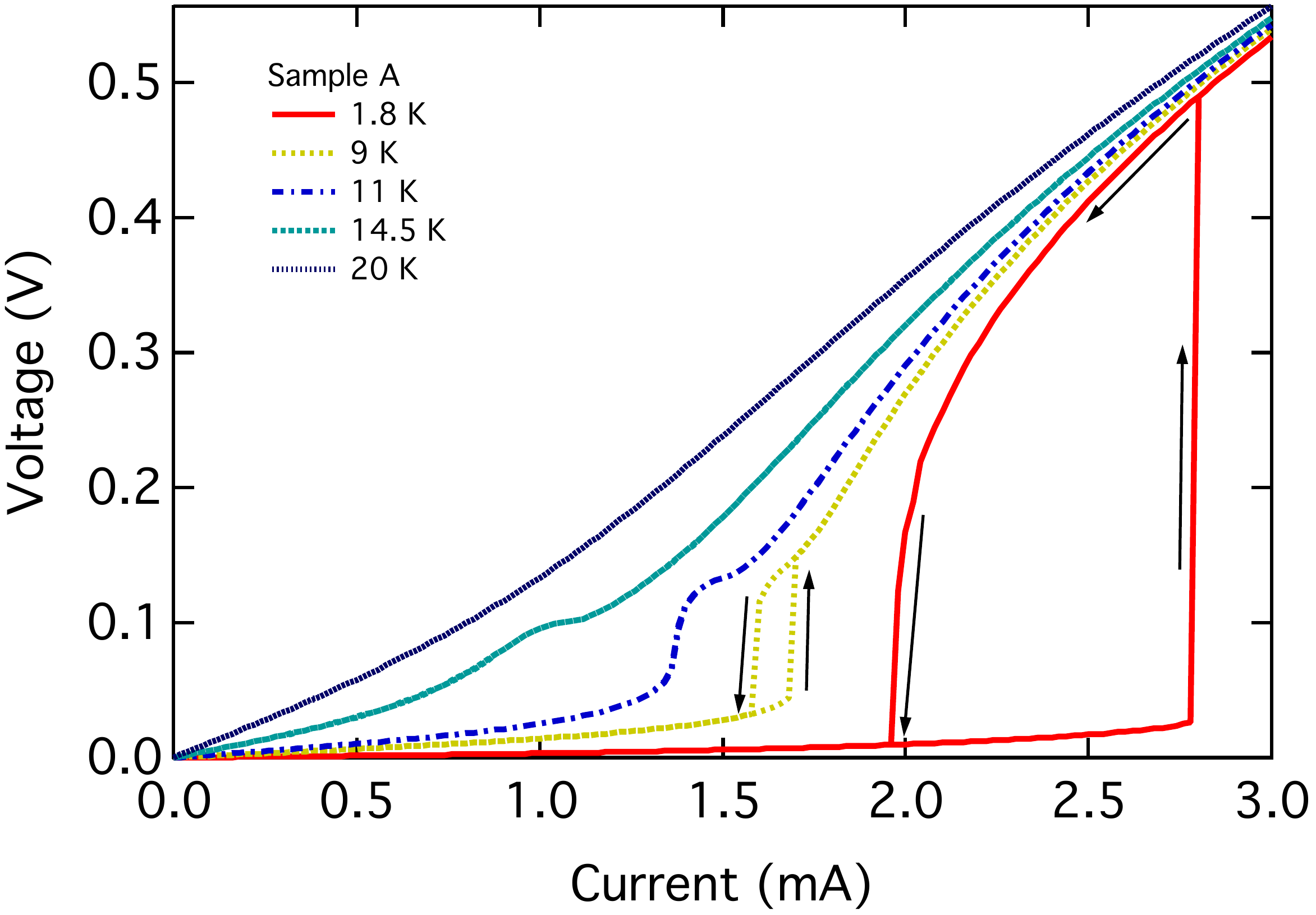}
\centering
\caption{\textbf{Current controlled measurements on sample A for temperatures above and below T$_{\rm{HO}}$.}  The up and down arrows denote increasing and decreasing current respectively.  Since the negative differential resistance observed in this measurement is such that the current is a single valued continuous function of voltage, the full IV character of the phase slips cannot be observed in the current-controlled regime.}
\label{fig:IControllA}
\end{figure}

\begin{figure}[h!]
\includegraphics[scale=0.5]{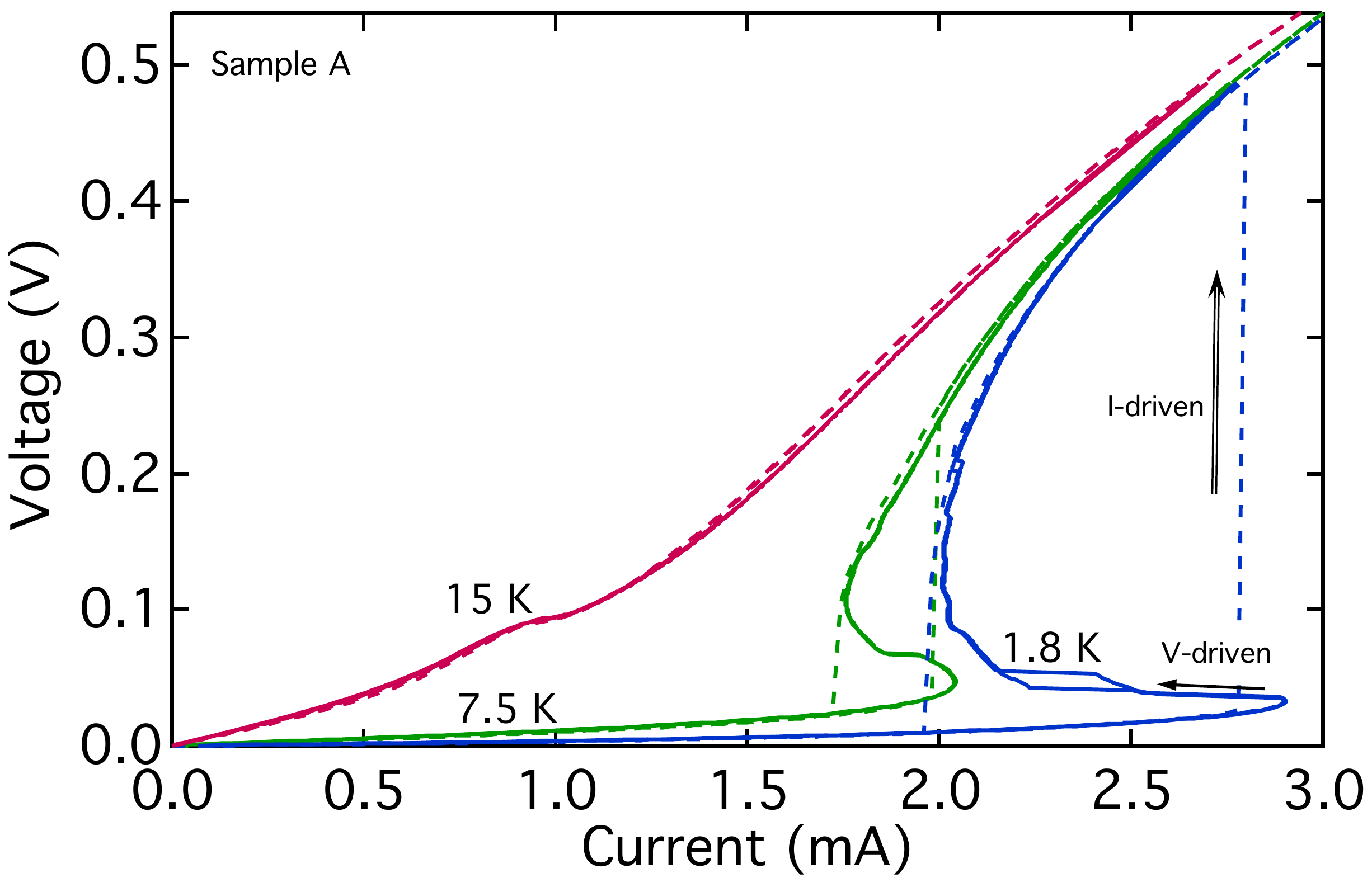}
  \centering
\caption{\textbf{Voltage-current characteristics of sample A at different temperatures.} For temperatures above T$_{\rm{HO}}$ the voltage driven measurements (solid lines) and current driven measurements (dotted lines) are in agreement.  For lower temperatures where phase slip behavior is observed, the current driven measurements fail to capture the full IV character.}
\label{fig:IcompareV}
\end{figure}

Four-terminal resistance measurements, including IV sweeps and differential conductance were performed on all three microstructured samples of \URuSi\ (labelled A, B, and C) to investigate the apparent phase slips in the hidden order state of the material.   Two-terminal measurements were also performed on samples A and B to calculate the contact resistance.  Given the ``N"-type negative differential resistance -- where current is a single valued  function of voltage, but voltage is a multivalued function of the current as shown in Figure~\ref{fig:NandS}(a) -- voltage controlled sweeps were used to capture the complete IV character of the phase slip behavior in these samples.  For completeness current controlled measurements, which give rise to ``S"-type negative differential resistance -- where voltage is a single valued function of current, but current is a multivalued function of the voltage as shown in Figure~\ref{fig:NandS}(b) -- were obtained on sample A at various temperatures, some of which are shown in Figure~\ref{fig:IControllA}, where arrows are used to denote changes in voltage due to increasing or decreasing current.  Graphing both current and voltage driven results together for representative temperatures of sample A, as shown in Figure~\ref{fig:IcompareV}, highlights the loss of information but otherwise accurate results obtained from current driven measurements.

\section{II. Electrical Transport Data for Samples B and C}
As reported in the main text for sample A, both samples B and C are measured in two different ways.  1) A continually sweeping voltage bias is applied along the sample emerged in superfluid $^4$He at 1.8 K for magnetic fields between 0 and 15 T.  2) A continually sweeping voltage bias is applied along the sample at zero magnetic field for constant temperatures both above and below the hidden order transition temperature, T$_{\rm{HO}}$.

\subsection{i. Sample B}

\begin{figure}[h!]
\includegraphics[scale=0.5]{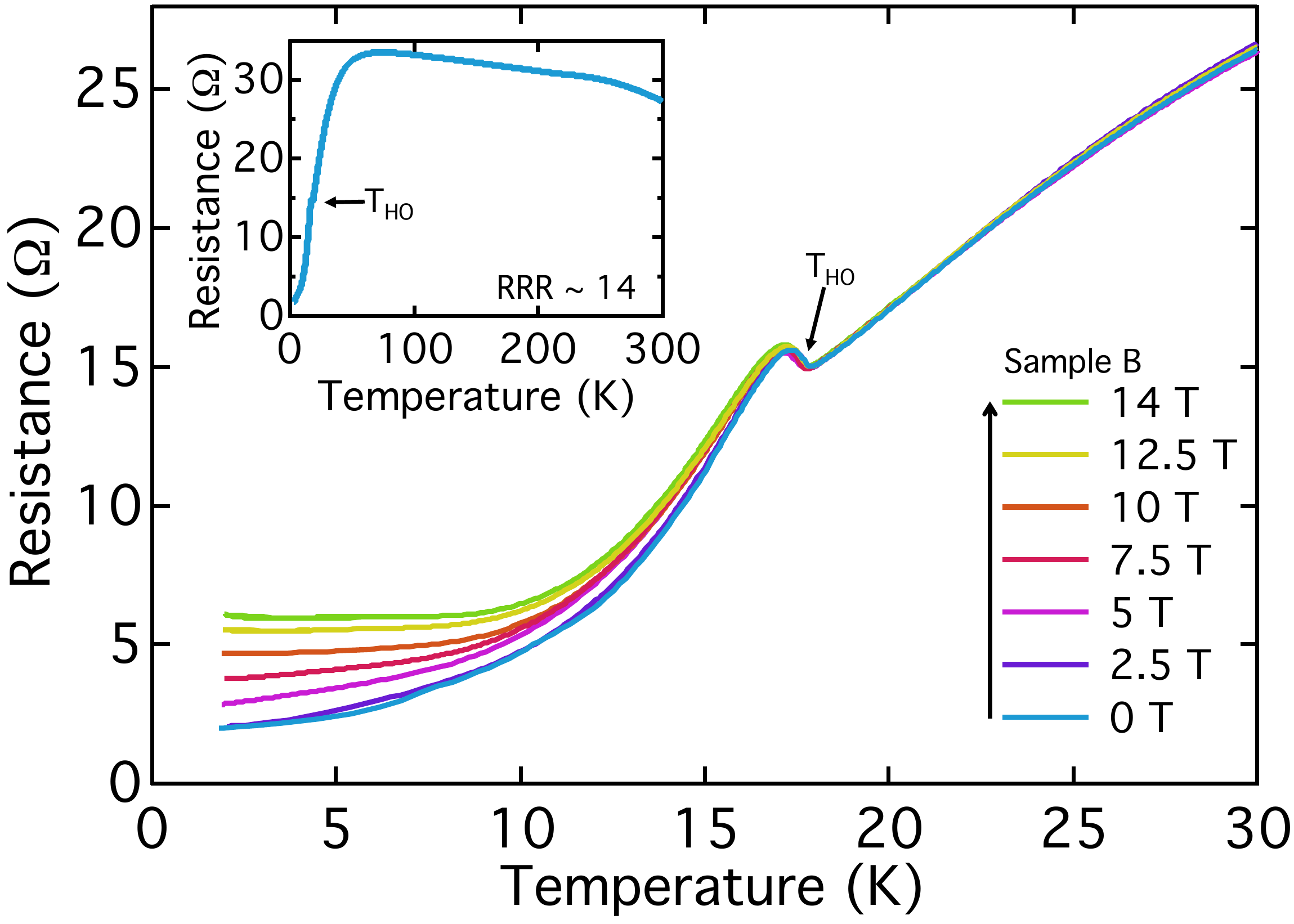}
   \centering
\caption{\textbf{Four-terminal resistance  for sample B as a function of temperature at constant magnetic fields up to 14 T.}   The inset shows the entire temperature range of the resistance at zero magnetic field from which the residual resistivity ratio (RRR) is determined.  T$_{\rm{HO}}$ denotes the temperature at the hidden order transition.}
\label{fig:SampleBRT}
\end{figure}

\begin{figure}[h!]
\includegraphics[scale=0.7]{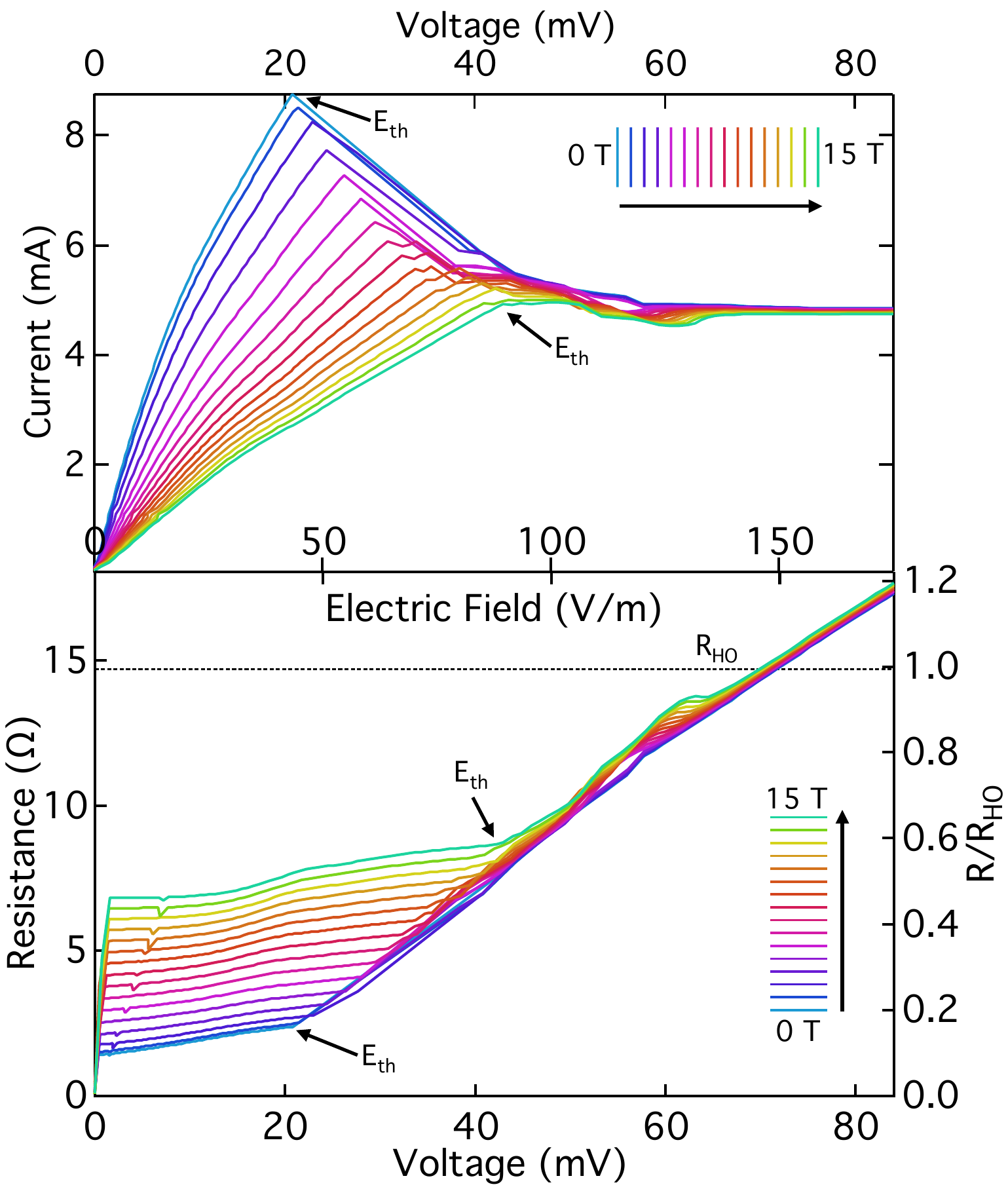}
  \centering
\caption{\textbf{Electrical transport properties of sample B obtained from voltage biased measurements at constant magnetic fields between 0 and 15 T.} (a) Current and (b) four-terminal resistance as a function of voltage and the resulting electric field, $E = V/d$, where \textit{d} is the effective length of the microstructured sample.   The right axis of the bottom figure corresponds to the ratio of the measured resistance (R) to the resistance at hidden order (R$_{\rm{HO}}$) in this sample.  At a threshold electric field, E$_{\rm{th}}$, a drop in current (increase in resistance) corresponds to a phase slip in the material, with additional signatures of the phase slip behavior observed between 100 - 150 V/m, that are consistent with variations in the electric field along the microstructured samples due to differences in the corner thickness of the meanders.}
\label{fig:SampleBIVR}
\end{figure}

Sample B has a microstructured pattern similar to A but with less meanders, resulting in a slightly lower resistance value at the base temperature of 1.8 K as noted in Table 1 of the main text.  Consequently, the resistance value at hidden order, R$_{\rm{HO}}$ and the residual resistivity ratio (RRR = $\rho_{\rm{300K}}/\rho{\rm_{1.8K}}$) for sample B are much smaller than sample A at 14$\Omega$ and 14.8 $\Omega$ respectively, as shown in Figure~\ref{fig:SampleBRT}.   Figure~\ref{fig:SampleBIVR}(a) plots the current through the microstructured sample as a function of voltage (and electric field) in superfluid $^4$He at different magnetic fields.   Qualitatively the observed behavior is similar to what was reported in the main text for A; at a threshold electric field the initial ohmic behavior abruptly changes with the onset of increased dissipation within the hidden order state.  However in this sample the transition to the more dissipative state is less pronounced than what was observed in sample A.  This is most notable at the highest magnetic fields where there is no sharp current maximum.   Additionally at higher magnetic fields effects of heating become increasingly apparent, as indicated by deviations from linear behavior before E$_{\rm{th}}$.  The heating effects are more discernible when the data is graphed as a function of resistance versus voltage as in Figure~\ref{fig:SampleBIVR}(b).  This heating is consistent with the sample resistance at 1.8 K for sample B being comparable to the contact resistance (R$_{\rm{contact}} \sim 5.1$~K) over the entire magnetic field range, which results in more Joule heating.   Despite this observed heating, the threshold electric field up to 15 T still occurs below R/R$_{\rm{HO}}$ = 1 (right axis, Figure~\ref{fig:SampleBIVR}(b)), indicating that the observed behavior is still due to phase slips and not heating through T$_{\rm{HO}}$.  

\begin{figure}[h!]
\includegraphics[scale=0.5]{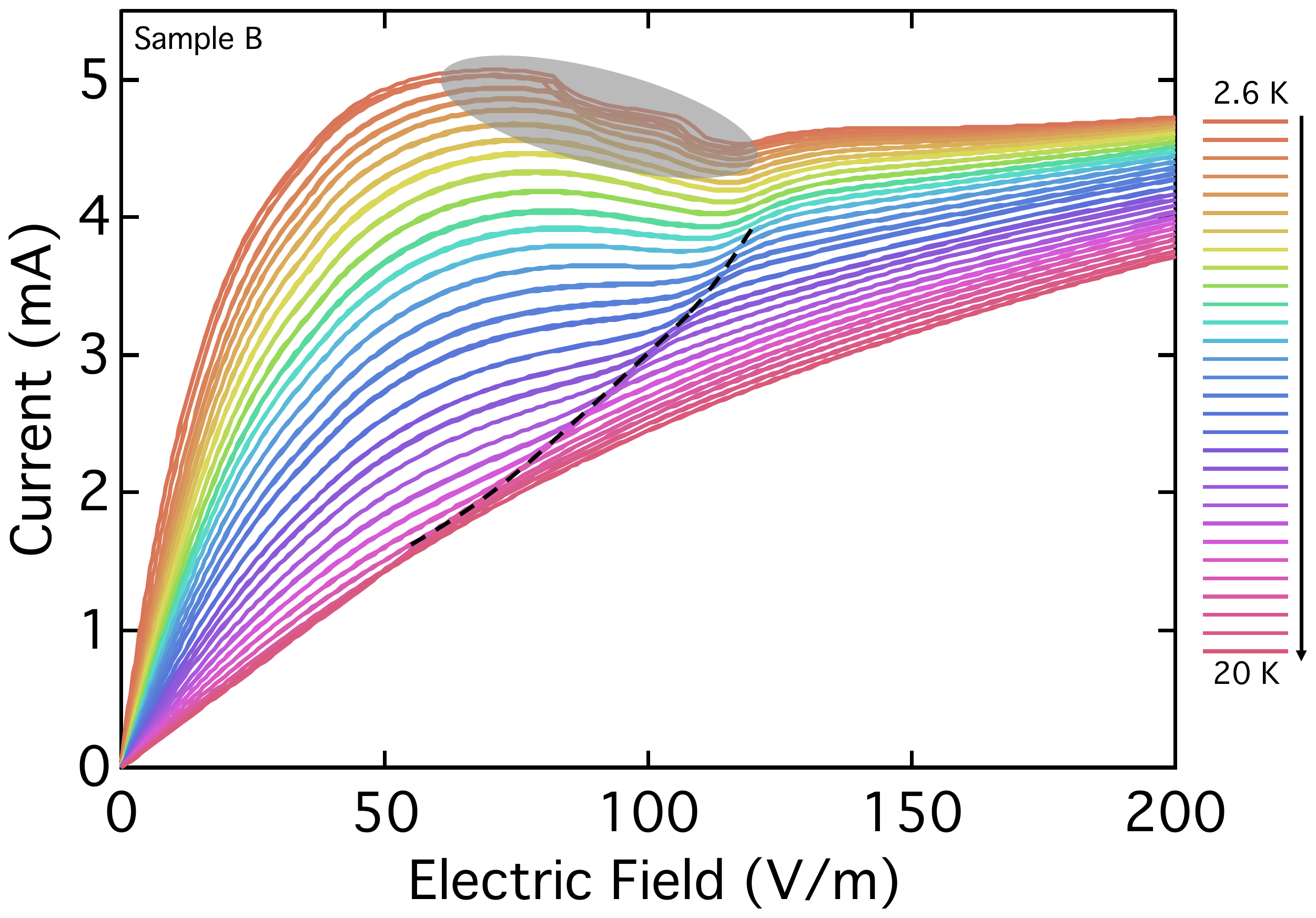}
   \centering
\caption{\textbf{Current as a function of the resulting electric field (due to an applied voltage) for sample B at temperatures between 2.6 K and 20 K.}  The shaded region highlights the negative differential conductance behavior that is associated with the observation of phase slips in the material, while the dotted line highlights positive differential conductance that is associated with the suppression of hidden order due to heating through the transition temperature.}
\label{fig:SampleBT}
\end{figure}

The temperature dependence of the IV character in sample B at elevated temperatures between 2.6 K and 20 K is shown in Figure~\ref{fig:SampleBT}.  As observed in sample A, a region of negative differential conductance (NDC, highlighted region) exists for the lowest measured temperatures and corresponds to the increased dissipation in the hidden order state associated with the development of phase slips in the material at E$_{\rm{th}}$.  The positive differential conductance behavior is observed for the lowest temperatures and is denoted by the black dotted line in Figure~\ref{fig:SampleBT}.  As was the case in sample A, this positive differential conductance behavior corresponds to the thermal suppression of hidden order and therefore bounds the negative differential conductance behavior firmly within the hidden order state.  However, unlike sample A, it appears as if T$_{\rm{HO}}$ is reached at all nominal temperatures due to heating from the applied voltage in sample B.

\subsection{ii. Sample C}
\begin{figure}[h!]
\includegraphics[scale=0.5]{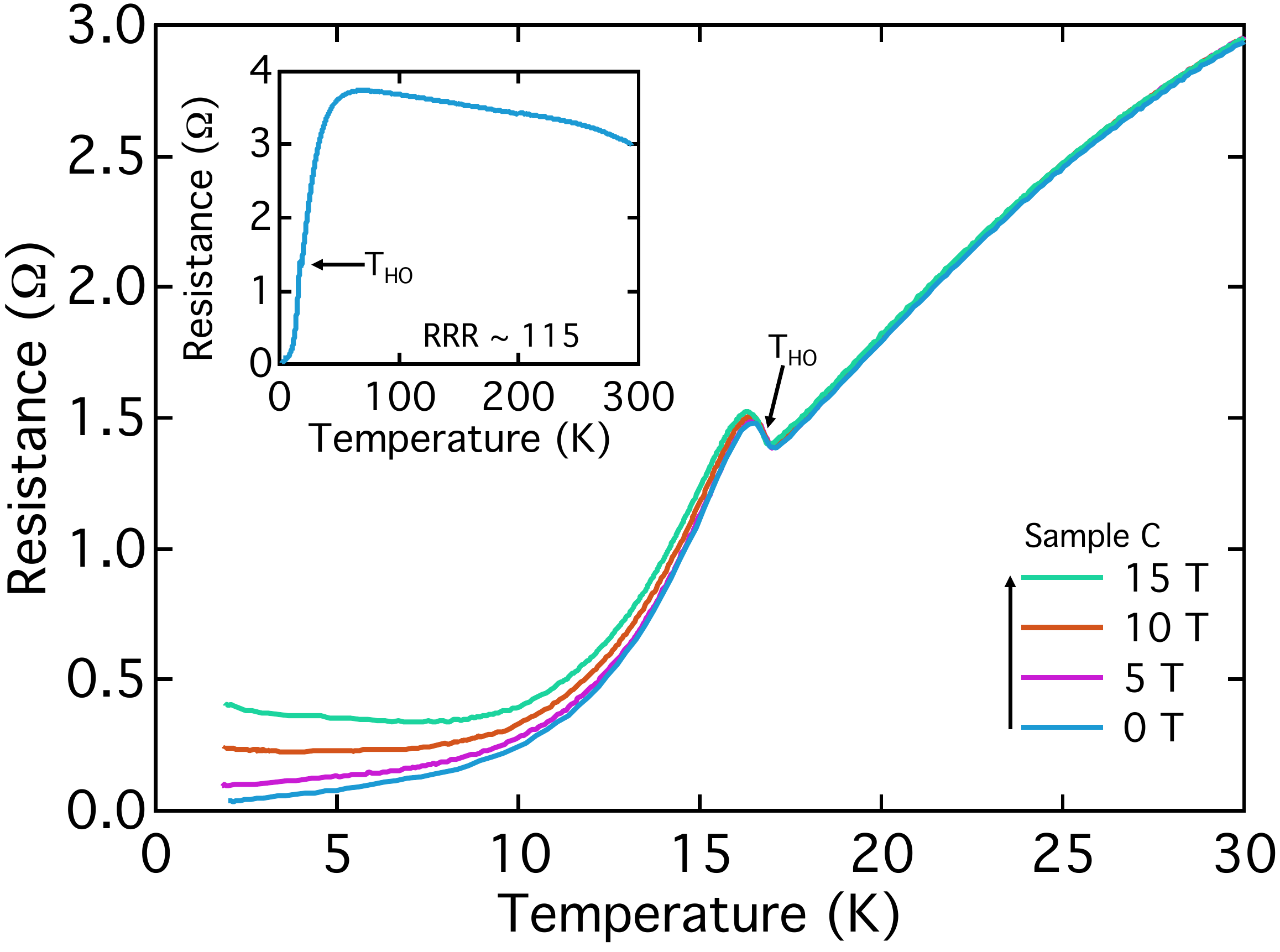}
   \centering
\caption{\textbf{Four-terminal resistance for sample C as a function of temperature at constant magnetic fields up to 15 T.}   The inset shows the entire temperature range of the resistance at zero magnetic field from which the residual resistivity ratio (RRR) is determined.  T$_{\rm{HO}}$ denotes the temperature at the hidden order transition.}
\label{fig:SampleCRT}
\end{figure}

The structure of sample C differs significantly from the other two samples examined in this study, as it consists of only one thin bar, an order of magnitude shorter than the other two samples. As a result, both the resistance at 1.8 K and at T$_{\rm{HO}}$ were significantly smaller for this sample at 20 m$\Omega$ and 1.47 $\Omega$ respectively.  In contrast to the small resistance values, the RRR of sample C is the largest of all three samples at 115, and the magnetoresistance at 1.8 K changes by an order of magnitude similar to that of sample A, as shown in Figures~\ref{fig:SampleCRT}.  

\begin{figure}[h!]
\includegraphics[scale=0.7]{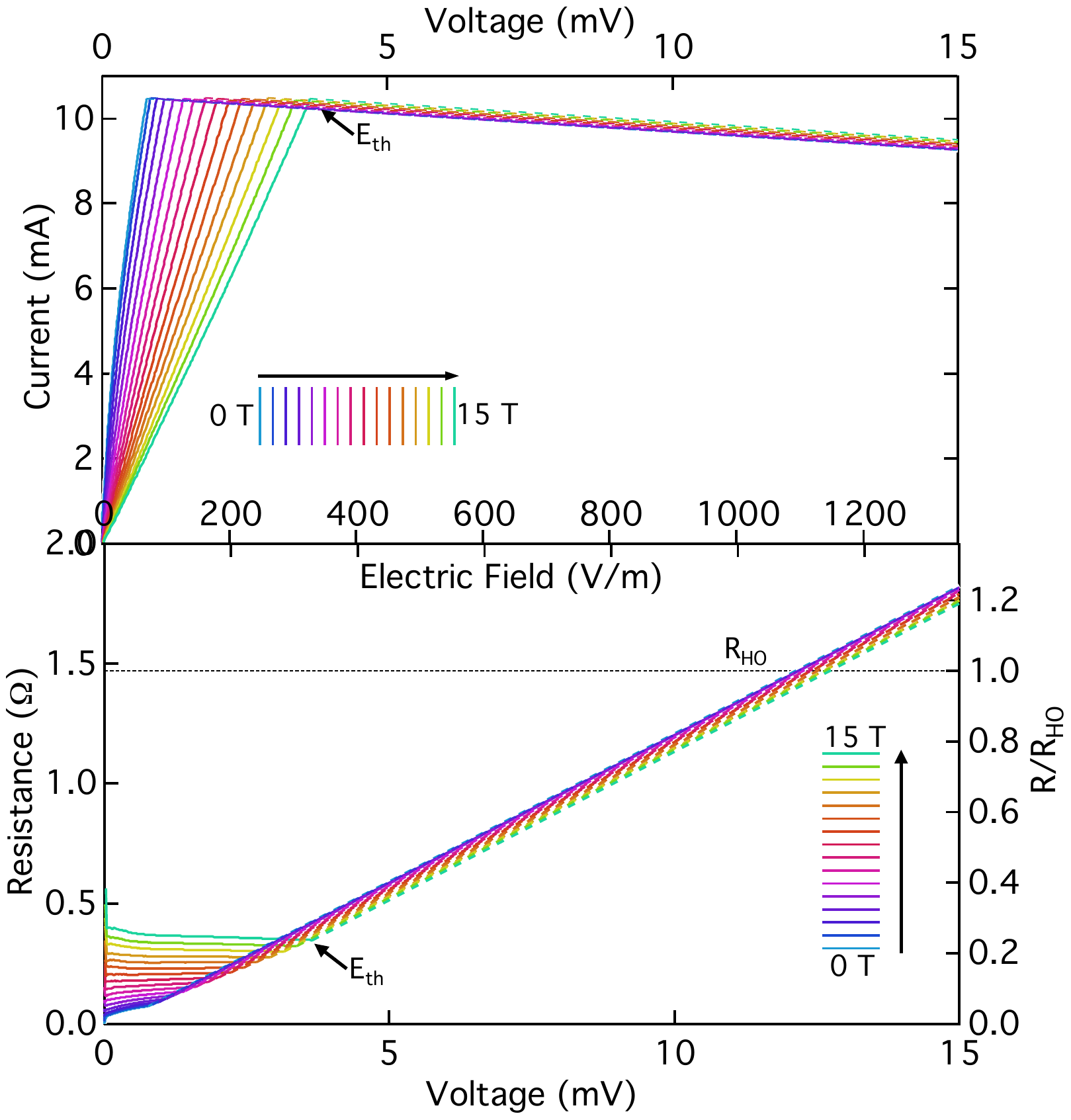}
   \centering
\caption{\textbf{Electrical transport properties of sample C obtained from voltage biased measurements at constant magnetic fields between 0 and 15 T.} (a) Current and (b) four-terminal resistance as a function of voltage and the resulting electric field, $E = V/d$, where \textit{d} is the effective length of the microstructured sample.  The right axis of the bottom figure corresponds to the ratio of the measured resistance (R) to the resistance at hidden order (R$_{\rm{HO}}$) in this sample. The low resistance value of the sample in comparison to the contact resistance gives rise to IV character that is more similar to current-controlled measurements; nevertheless, representative threshold electric fields, E$_{\rm{th}}$ are labelled accordingly.}
\label{fig:SampleCIVR}
\end{figure}

Unlike the two previously discussed samples, voltage controlled measurements on sample C were complicated by a contact resistance two orders of magnitude larger than the sample's resistance at 0 T and 1.8 K.  Consequently, the contacts act as the primary source of dissipation, resulting in significant Joule heating, and eliminating the ability to control the electric field along the sample -- as opposed to the current -- via the applied voltage.  This lack of voltage control in sample C measured in superfluid $^4$He at 1.8 K at different magnetic fields results in an IV character that is qualitatively similar to current control measurements (Figure~\ref{fig:NandS}(b)), as shown in Figure~\ref{fig:SampleCIVR}.  Nevertheless, the electric field at which the IV discontinuity occurs is consistent with the other devices despite the very different geometries.  It is of interest to note that the current values at E$_{\rm{th}}$ appear current limited, i.e. the maximum current is the sample for all E$_{\rm{th}}$(B) and as a result maximum current densities for this sample as a function of magnetic field are not included in Figure 3(d) of the main text.  As with the other samples, the magnetic field dependence of the IV character for sample C can also be graphed as resistance versus voltage (and electric field) as shown in Figure~\ref{fig:SampleCIVR}(b), however the amount of information gained from examining the data this way is limited given the large range over which no values are recorded.  

\begin{figure}[h!]
\includegraphics[scale=0.5]{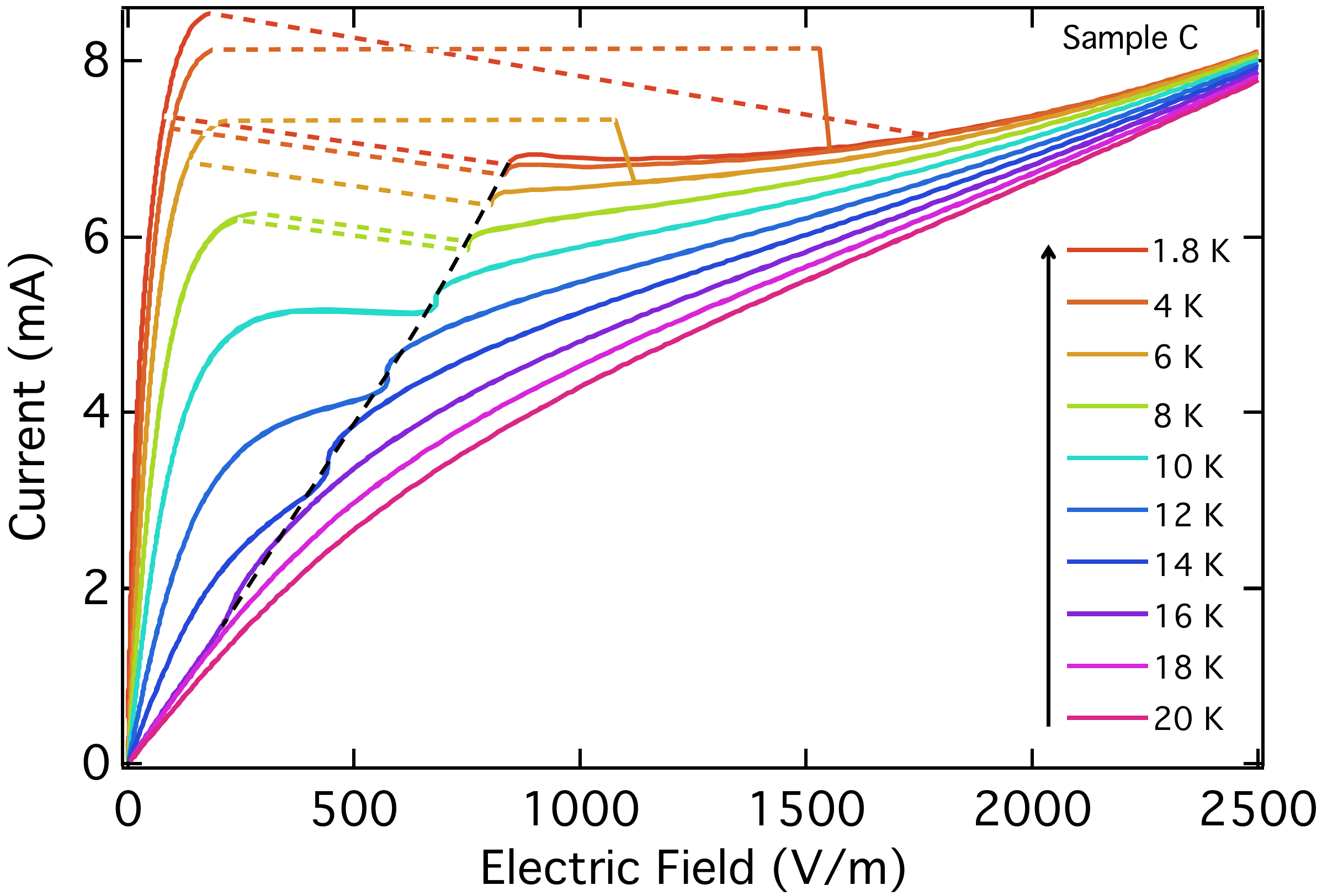}
   \centering
\caption{\textbf{Current as a function of the resulting electric field (due to an applied voltage) for sample C between 1.8 K and 20 K.}  The current controlled behavior in this sample is revealed in the discontinuity of the data at low temperatures, given by the colored dotted lines, that obscure the full IV character of the phase slips.  The black dotted line emphasizes the suppression of hidden order due to heating through the transition temperature.  }
\label{fig:SampleCT}
\end{figure}

The temperature dependence of sample C's IV character appears ``S"-type in so much that there are large discontinuities in the data at the lowest temperatures, shown as dotted lines in Figure~\ref{fig:SampleCT}, which hide the complete signature of the phase slips.  Joule heating is also observed over a large range of elevated temperatures, where the suppression of hidden order due to heating through T$_{\rm{HO}}$ corresponds to the small increase in current as emphasized by the black dotted line in the figure.  Nevertheless, the estimated threshold electric fields and maximum current densities for sample C show qualitatively similar temperature dependence as observed in samples A and B, further supporting the intrinsic nature of the reported phase slip behavior even in samples with very different geometries and RRR values.  

\section{III. Energy and Length Scales}
Although we have ruled out the possibility of the negative differential conductance behavior reflecting the suppression of hidden order due to the application of an electric field it is still informative to look at the energetics of the applied electric field on \URuSi\ in the context of both purely local and itinerant electron behavior.   Since electric field does not intrinsically provide an energy scale, it is necessary to look at the length scale over which it acts.  If we consider the \textit{f}-electrons as itinerant, then one can think about the effect of an electric field as a displacement of the Fermi surface away from its equilibrium position in the band structure. This acceleration of charge is balanced by scattering events that typically prevent large Fermi surface displacement and a metallic crystal from entering the regime of Bloch oscillations \cite{Grondin85}.  While never observed in normal 3D metals, artificial superlattice structures can be engineered such that the carriers can be accelerated past the point of the band inversion, in which case the Bloch oscillator regime presents as a similar region of NDC, whereby current decreases with increasing voltage \cite{Esaki70}.   To estimate the Fermi surface displacement a scattering rate $\tau$ for each pocket of the Fermi surface is estimated using quantum oscillation measurements performed on similar residual resistivity ratio (RRR) samples \cite{Altarawneh11, Harrison13}, as the magnetic field at which the onset of quantum oscillations occurs is proportional to the cyclotron frequency ($\omega_{\rm{C}}\tau \approx1 $).  Furthermore, the temperature dependence of the oscillation amplitude provides an independent measure of the effective mass, $m^*$, and hence the cyclotron frequency, $\omega_{\rm{C}} = \frac{eB}{m^*}$, where $e$ is the electronic charge and $B$ is the applied magnetic field.  Based on our experimental results that provide a threshold electric field on the order of 100 V/m, in combination with a relaxation time on the order of 5 ps, the estimated Fermi surface displacement, $\Delta k = \frac{eE\tau}{\hbar}$, ends up being four orders of magnitude smaller than the zone boundary which corresponds to the approximately 4~\AA\ unit cell.   

One can also consider the effect of electric field upon the local {\it f-}electron energetics. When acting on an atomic scale ($\sim$ 1 \AA) the observed threshold electric field is a perturbation five orders of magnitude smaller (10~neV) than the hidden order energy scale of ~2~meV and relevant spin-orbit and crystal-field energies \cite{Mydosh11}. Comparing both purely local and itinerant behavior suggests that neither picture is sufficient to explain the significant effect of the observed threshold electric field, suggesting that it must be acting upon a length on the order of tens-of-thousands of unit cells, i.e. one-to-tens of micrometers, a  length comparable to the lateral device dimensions.  This length scale is also compatible with the observation of phase slips in the devices. Since phase slips in 1D superconductors are only observed when the transverse dimensions are less than $\xi$, it implies that the width of the devices, $\sim 1\ \mu$m, are smaller than the coherence length of the hidden order wavefunction and that the coherence length approaches the single meander length of the two larger devices, $\sim 70\ \mu$m.

\end{document}